\documentclass[apj]{emulateapj}
\usepackage{natbib}
\usepackage{apjfonts}
\usepackage{ulem}
\usepackage{graphicx}
\usepackage{amsfonts}
\usepackage{amsmath}
\usepackage{amssymb}
\usepackage{color}
\input{colordvi.tex}
\usepackage{bm}

\usepackage{natbib}
\bibpunct{(}{)}{;}{a}{}{,} 

\begin{document}
\title{CO Component Estimation Based on the Independent Component Analysis}
\author{
Kiyotomo Ichiki \altaffilmark{1},
Ryohei Kaji \altaffilmark{2},
Hiroaki Yamamoto \altaffilmark{2},
Tsutomu T. Takeuchi \altaffilmark{2}, Yasuo Fukui \altaffilmark{2}
}
\altaffiltext{1}{%
Kobayashi-Maskawa Institute for the Origin of Particles and the Universe, Nagoya University, Nagoya
464-8602, Japan}
\altaffiltext{2}{%
Department of Physics and Astrophysics, Nagoya University, Nagoya
464-8602, Japan
}

\begin{abstract}{
Fast Independent Component Analysis (FastICA) is a component separation
 algorithm based on the levels of non-Gaussianity. Here we apply the
 FastICA to the component separation problem of the microwave background
 including carbon monoxide (CO) line emissions that are found to
 contaminate the PLANCK High Frequency Instrument (HFI)
 data. Specifically we prepare 100GHz, 143GHz, and 217GHz mock microwave
 sky maps including galactic thermal dust, NANTEN CO line, and the
 Cosmic Microwave Background (CMB) emissions, and then estimate the
 independent components based on the kurtosis.  We find that the FastICA
 can successfully estimate the CO component as the first independent
 component in our deflection algorithm as its distribution has the
 largest degree of non-Gaussianity among the components. By subtracting
 the CO and the dust components from the original sky maps, we will be
 able to make an unbiased estimate of the cosmological CMB angular power
 spectrum.}
\end{abstract}


\section{INTRODUCTION}

Precise measurements of the Cosmic Microwave Background (CMB)
anisotropies have been a powerful probe into the early universe and
cosmology. Experiments such as COBE \citep{1996ApJ...464L...1B},
BOOMERanG \citep{2006ApJ...647..799M}, WMAP \citep{2011ApJS..192...18K}
have already placed strong constraints on the parameters of the
cosmological model, such as the age of the universe, baryon and cold
dark matter densities and so on.  The third generation CMB satellite,
PLANCK \citep{2010A&A...520A...1T}, is expected to release its
cosmological results soon which will include constraints on the
amplitude of primordial gravitational waves, the spectral index and its
running of the primordial curvature perturbations, the amount of
primordial non-Gaussianity and so on, and thereby will give constraints
on physics of the early universe such as inflation.

Such cosmological information can be obtained only when sources of
uncertainty are removed successfully. With recent high-resolution and
sensitive instruments, the main source of uncertainty is the
contamination by foreground emissions from the Galaxy, rather than the
instrumental noise itself. Therefore component separation methods have
been progressively developed so far based on the analyses of data at
different frequencies and different dependencies on frequency of the
astrophysical emission laws (for a recent review, see
\cite{2009AIPC.1141..222D}). The methods include template fitting
\citep{2009MNRAS.397.1355E,2011ApJ...737...78K}, Internal Linear
Combination \citep{2004ApJ...612..633E}, Correlated Component Analysis
\citep{2006MNRAS.373..271B}, Maximum Entropy Method
\citep{1998MNRAS.300....1H} and so on, where the differences are in the
way how they model the data and the assumptions made on the foreground
components.

Along with the synchrotron and thermal dust emissions which constitute
substantial portion of the foreground component of the Galaxy at
microwave frequencies, it now becomes clear that the rotational
transitions of carbon monoxide (CO) significantly contaminate the PLANCK
observing bands \citep{2011A&A...536A...6P}.  In particular, the
frequencies of the lowest two rotational transitions of CO, namely
J=(1-0) and J=(2-1), are at the first and third transmission bands of
PLANCK's High Frequency Instrument (HFI), that is, at $100$ and $217$ GHz
bands. Therefore we must develop a method to remove this contribution
for cosmological analysis.

A simple way to remove such contribution may be to use a template with help
from the other CO line surveys, such as Columbia survey (e.g.,
\cite{2001ApJ...547..792D}) and NANTEN Galactic Plane Survey (NGPS) survey by NANTEN telescope
\citep{2001PASJ...53.1017O,2004ASPC..317...59M}. However, these surveys are dedicated mainly
to the Galactic plane where the most of the molecular clouds has been
found and there has been no full sky CO map to be compared with the
PLANCK full sky data. Therefore it will be helpful to find a method to
obtain (even rough) information about the distribution of the CO
molecular clouds, especially at high galactic latitudes, from the PLANCK
data alone. 

To this end we consider a fast component separation method to extract
the CO distribution based on the Independent Component Analysis
(FastICA) \citep{Hyvarinen97afast}. The FastICA has several advantages
in comparison with the methods mentioned above; among them the most
important point is that FastICA method needs no prior assumptions about
the distribution of the foreground components and their frequency
dependences (therefore it is called as Blind Source Separation in the
statistics community). Instead the method uses statistical independency
to separate the components as described below shortly. Therefore the
method may be suitable to estimate the distribution at the current
moment when we do not know much about the CO distribution on the full
sky and its relative contributions to the PLANCK observing
bands. Several applications of the FastICA to the CMB component
separation problem can be found in the literature, which includes
applications to the COBE
\citep{2002MNRAS.334...53M,2003MNRAS.344..544M}, BEAST
\citep{2006MNRAS.369..441D} WMAP data
\citep{2007MNRAS.374.1207M,2010MNRAS.402..207B} and simulated 21 cm maps
\citep{2012MNRAS.423.2518C}. Here we first apply the method to extract
the CO component assuming the PLANCK HFI and evaluate its performance by
means of Monte Carlo simulations, in terms of the angular power spectrum
of temperature anisotropies which is one of the most important statistic
for cosmological analysis.

\section{FastICA method}

\begin{figure}
\begin{minipage}[m]{0.5\textwidth}
\rotatebox{0}{\includegraphics[width=1.0\textwidth]{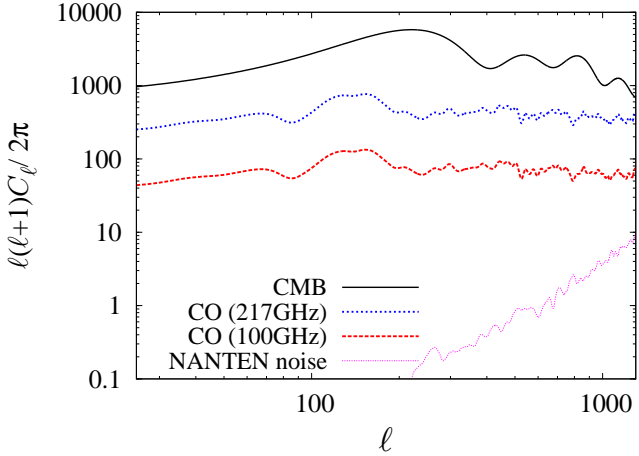}}
\end{minipage}
\caption{Angular power spectra of CO line emissions at the 100GHz band in
 the MBM  and Pegasus region 
 observed by the NANTEN telescope with estimated error magnitude (red
 dashed line). 217GHz
 signal is estimated assuming the line ratio of
 \cite{2000ApJ...535..211I} (blue dotted line). CO
 angular power spectrum gives more significant contribution at smaller angular
 scales relative to the cosmological signal (black line). Thanks to the high
 resolution of the NANTEN telescope the noise 
 level is negligible in our study (magenta dash-dotted line).  } \label{fig:Cls_COnoise}
\end{figure}

ICA assumes that observed maps are given by a superposition of
independent astrophysical components and the cosmological CMB. The ICA
model is given by
\begin{equation}
T^{j}(\hat{n}_i)=M^j_k S^k(\hat{n}_i)~,
\end{equation}
where $T^{j}(\hat{n_i})$ are the observed temperatures with the $j$-th
band at the sky direction $\hat{n_i}$, $M^j_k$ is the mixing matrix, and
$S^{k}$ with $k=1,2,3$ are the three independent sources 
considered in this paper that correspond to CO, thermal dust and CMB emissions, respectively. In our
simulation  
we consider PLANCK's $j=100$, $143$ and $217$ GHz bands.  The ICA algorithm
estimates the sources and the mixing matrix simultaneously by maximizing
the degree of non-Gaussianity of the variable $Y^{k}$ with the matrix
$W^k_j$,
\begin{equation}
Y^{k}(\hat{n}_i) = W^k_j T^j(\hat{n}_i)~.
\end{equation}
Naively, this process is equivalent to maximizing ``independency''
between the variables $Y^{k}$ because of the central limit theorem.  When
the non-Gaussianity of $Y^k$ takes the maximum, $W$ should be ${M^{-1}}$
and $Y^k$  approaches $S^k$. In the present analysis we consider a noisy
ICA model which is given by 
\begin{equation}
T^{j}(\hat{n}_i)=M^j_k
S^k(\hat{n}_i)+N^j(\hat{n}_i), 
\label{eq:3}
\end{equation}
where instrumental (white) noise terms
$N^j$ are taken from the PLANCK specification
\citep{2011A&A...536A...1P}. Because the noise level of NANTEN telescope
is not significant compared to that of the PLANCK as shown in
Fig. \ref{fig:Cls_COnoise}, so we have neglected it.

In what follows, we use the vector notation and write $\bm{T}$ instead
of $T^j$, etc., for clarity. Following the standard ICA procedure, we
first quasi-whiten the observed data. The whitening is done by the
operations \citep{1999ISPL....6..145H}
\begin{eqnarray}
\bm{x}(\hat{n}_i) &=& (\bm{T}(\hat{n}_i)-\bm{\bar{T}})~, \\
\bm{\tilde{x}}(\hat{n}_i) &=& {(\bm{C}-\bm{\Sigma})^{-\frac{1}{2}}}
 \bm{x}(\hat{n}_i) ~,\\
\bm{\tilde{\Sigma}}&=&{(\bm{C}-\bm{\Sigma})^{-\frac{1}{2}}} \bm{\Sigma} {(\bm{C}-\bm{\Sigma})^{-\frac{1}{2}}}~,
\end{eqnarray}
where $\bm{\bar{T}}=E\{\bm{T}\}$ is the mean of temperature in each
observed band, $\bm{C}=E\{{\bm{x}\bm{x}^T}\}$ is the covariance matrix
of the observed data, $\bm{\Sigma}=E\{\bm{N}\bm{N}^T\}$ is the known
noise covariance matrix, and $\bm{\tilde{\Sigma}}$ is that after
quasi-whitening. The ensemble average $E\{\}$ is estimated from the
sample average from the observed pixels $\hat{n}_i$. Thus, the problem
is recasted to finding a matrix $\bm{W}$ which maximizes the levels of
non-Gaussianity of the variables $\bm{y}\equiv \bm{W} \bm{\tilde{x}}$.

In the analysis we adopt the deflation algorithm, i.e., we estimate the
independent components one-by-one by maximizing the non-Gaussianity of
the variable $y(\hat{n}_i)=\bm{w}^T \bm{\tilde{x}}(\hat{n}_i)$, under a
constraint $|\bm{w}|^2=1$. In
this case, the vector $\bm{w}^T$ is a row of the matrix $\bm{W}$. 
To find $\bm{w}^T$ which maximizes the non-Gaussianity of the
variable $y$, we need an evaluation function of the level of
non-Gaussianity. In the present analysis we use the kurtosis as the
evaluation function $g(y)$:
\begin{equation}
g(y)={\rm kurt}(y) = E\{y^4\}-3\left(E\{y^2\}\right)^2~.
\end{equation}
The function $g(y)$ takes the minimum $g(y)=0$ when the variable $y$ is
Gaussian distributed.  The gradient of the kurtosis is given by
\begin{equation}
\frac{\partial g}{\partial \bm{w}}=4E\{(\bm{w}^T\bm{\tilde{x}})^3
 \bm{\tilde{x}}\} -12 \bm{w}^T\left(\bm{I}+\bm{\tilde{\Sigma}}\right)
\bm{w}\left(\bm{I}+\bm{\tilde{\Sigma}}\right)\bm{w}~,
\end{equation}
where we have used the fact that $E\left\{\bm{\tilde{x}}
\bm{\tilde{x}}^T\right\}=\bm{I}+\bm{\tilde{\Sigma}}$~. In the standard
gradient method, the parameter $\bm{w}$ should be updated by
\begin{equation}
\Delta\bm{w}\propto \frac{\partial g}{\partial \bm{w}} 
 \propto E\{(\bm{w}^T\bm{\tilde{x}})^3
 \bm{\tilde{x}}\} -3 \bm{w}^T\left(\bm{I}+\bm{\tilde{\Sigma}}\right)
\bm{w}\left(\bm{I}+\bm{\tilde{\Sigma}}\right)\bm{w}~.
\end{equation}
 Because we have restricted the parameter space by the constraint
 $|\bm{w}|^2=1$, the vector $\bm{w}$ should satisfy the condition
 $\bm{w}\propto \Delta\bm{w}$ at the stable point. Therefore one obtains
 the fixed point algorithm \citep{Hyvarinen97afast},
\begin{equation}
 \bm{w}_{\rm new} = E\{(\bm{w}^T\bm{\tilde{x}})^3
 \bm{\tilde{x}}\} -3 \bm{w}^T\left(\bm{I}+\bm{\tilde{\Sigma}}\right)
\bm{w}\left(\bm{I}+\bm{\tilde{\Sigma}}\right)\bm{w}~,
\end{equation}
which is followed by a normalization $\bm{w}_{\rm new}\leftarrow
\bm{w}_{\rm new}/|\bm{w}_{\rm new}|$. The above procedure maximizes the
non-Gaussianity of the variable $y=\bm{w}^T \bm{\tilde{x}}$ in terms of
the kurtosis.

When we need to estimate more than one independent component, we can
repeat the above procedure. In this case, an orthogonalization step must
be operated at every iteration before the normalization step to prevent
the different vectors $\bm{w}^{(n)}$ from converging to the same vector,
where $(n)$ means the $n$-th independent component. The
orthogonalization can be done, for example, by Gram-Schmidt-like
decorrelation method:
\begin{equation}
\bm{w}^{(p+1)}=\bm{w}^{(p+1)}-\Sigma_{j=1}^p (\bm{w}^{(p+1)} \cdot
 \bm{w}^{(j)}) \bm{w}^{(j)}~.
\end{equation}

\section{Sky Model}
\begin{figure}
\begin{minipage}[m]{0.5\textwidth}
\rotatebox{90}{\includegraphics[width=1.0\textwidth]{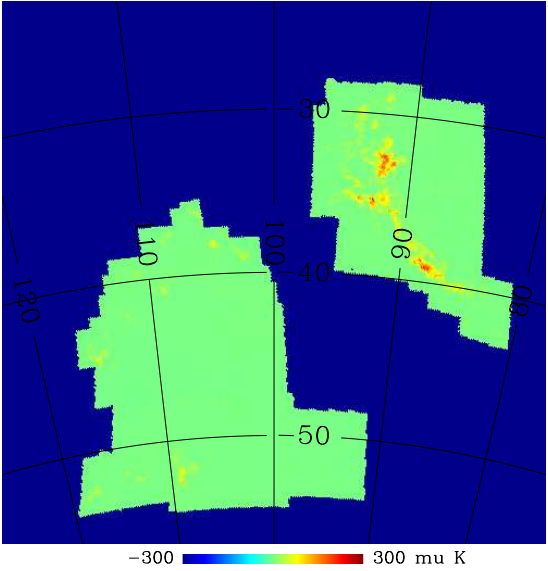}}
\end{minipage}
\caption{CO line emission intensity at the MBM  and Pegasus region observed by NANTEN
 telescope.  The unobserved (masked) region is shown in blue.} 
\label{fig:MBM_region}
\end{figure}

\begin{figure}
\begin{minipage}[m]{0.5\textwidth}
\centering
\rotatebox{90}{\includegraphics[width=0.9\textwidth]{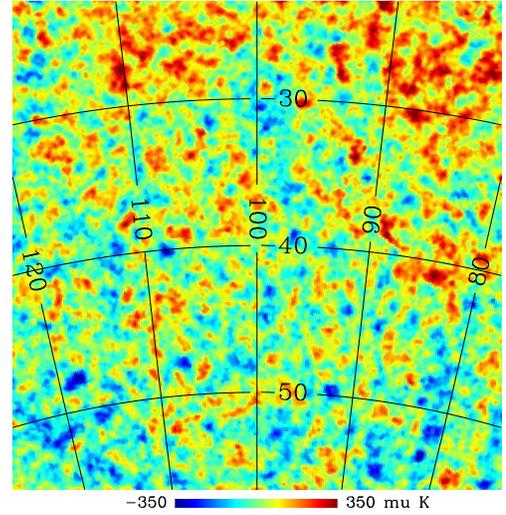}}
\end{minipage}
\begin{minipage}[m]{0.5\textwidth}
\vspace*{-0.7cm}
\centering
\rotatebox{90}{\includegraphics[width=0.9\textwidth]{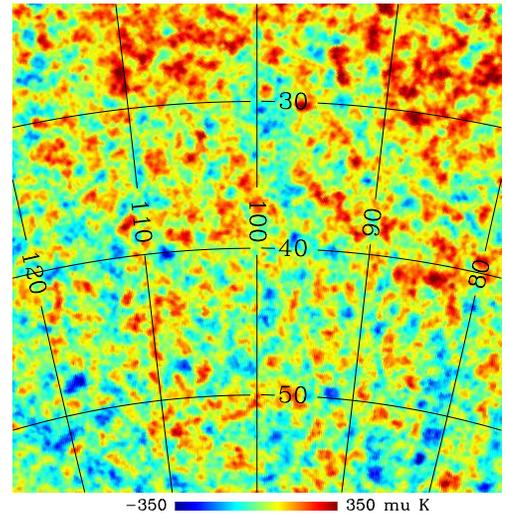}}
\end{minipage}
\begin{minipage}[m]{0.5\textwidth}
\vspace*{-0.7cm}
\centering
\rotatebox{90}{\includegraphics[width=0.9\textwidth]{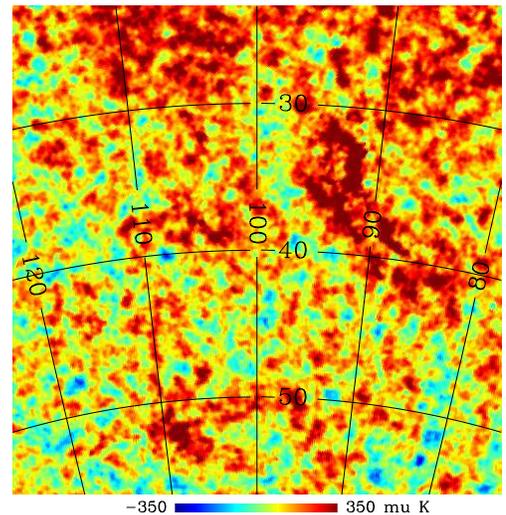}}
\end{minipage}
\caption{
Simulated CMB sky maps at 100GHz (top), 143GHz (middle), and 217GHz (bottom).
}
\label{fig:fig1}
\end{figure}

In Fig. \ref{fig:fig1} we show simulated sky maps at the $100, 143$, and
$217$ GHz PLANCK bands including the CMB, foreground components and the
PLANCK instrumental noises ($T^j$ in Eq.(1)). For the CMB component we
generate random Gaussian skies with the angular power spectrum
consistent with the WMAP 7 year results \citep{2011ApJS..192...16L}. The
skies are convolved with a spherical beam with the largest beam width
among the three bands, namely, ${\rm FWHM}=9.37$ arc-min at the $100$ GHz
band \citep{2011A&A...536A...1P}. For the instrumental noises we assume
white noises for about $14$ months observation with the amplitudes
\begin{eqnarray}
\Delta T [\mu K / {\rm pix}]
&=&\Delta T [\mu K / {\rm beam}] 
\frac{\theta_{\rm beam}}{\theta_{\rm pix}} \nonumber \\
&=&\left\{ \begin{array}{ll}
2.80 & (100{\rm GHz})~ \\
1.79 & (143{\rm GHz})~ \\
2.55 & (217{\rm GHz})~, \\
\end{array} \right.
\label{eq:noise}
\end{eqnarray} 
where $\theta_{\rm pix}=6.87$ arc-min (Gaussian) with the HEALPIX
parameter $N_{\rm side}=512$ \citep{2005ApJ...622..759G}. The noise
terms $N^j$ in Eq.(\ref{eq:3}) are realized randomly from normal
distributions with variances given by Eq. (\ref{eq:noise}).

For the foreground components we assume galactic thermal dust and CO
line emissions. In particular, thermal dust emissions and CO line
contaminations are prominent at $217$ GHz and the $100$ and $217$ GHz
bands, respectively.  For the thermal dust emissions we follow the
``model 8'' of Finkbeiner et al. \citep{1999ApJ...524..867F}, which
gives predictions of dust maps at microwave frequencies through
extrapolations from \cite{1998ApJ...500..525S}. The same model maps are
implemented in a recent paper \citep{2010ApJ...709..920S}, and slightly
different maps (``model 7'') are used in the PLANCK Sky Model
\citep{2012arXiv1207.3675D}. For the CO line contamination at the $100$
GHz band we use real data at MBM and Pegasus region observed by NANTEN
telescope \citep{2003ApJ...592..217Y,2006ApJ...642..307Y} and convert the NANTEN
velocity-integrated intensity map to the CMB temperature map at $100$
GHz band by multiplying the conversion factor $\alpha^{J=1-0}\equiv
T^{100\rm GHz}_{\rm CO}/I_{\rm CO}^{J=1-0}=14.2$ found by the PLANCK
team \citep{2011A&A...536A...6P}.  The FWHM of the NANTEN beam is about
$2.6$ arcmin and we smooth the map to $9.37$ arcmin by using the
subroutine {\it alteralm} in the HEALPix facilities in order to match
the FWHM of the PLANCK 100GHz band. While the CO line contamination at
the $143$ GHz band is found not to be significant
\citep{2011A&A...536A...6P} we must taken into consideration the
contamination at the $217$ GHz band where the transition J=(2-1) comes
in.  Because NANTEN data are not available for this
transition we make a simple assumption that the intensity of J=(2-1)
transition is proportional to that of J=(1-0).

Specifically, we make a toy sky map for the CO line emission at $217$
GHz by
\begin{equation}
T_{\rm CO}^{217{\rm GHz}}(\hat{n}_i) = R^{J_{2-1}}_{J_{1-0}} 
 \frac{\alpha^{J_{1-0}}}{\alpha^{J_{2-1}}} T_{\rm CO}^{100{\rm GHz}}(\hat{n}_i)~.
\end{equation}
Here the integrated line ratio, $R^{J_{2-1}}_{J_{1-0}}=0.77\pm 0.24$, is
taken from \cite{2000ApJ...535..211I}, in which they estimated the ratio
between J=(4-3), (2-1) and (1-0) at high galactic molecular clouds based
on the observations using the Antarctic Sub-millimeter Telescope, 
Remote Observatory and the Five College Radio Astronomy Observatory.

Because the CO line emission data is limited to the MBM  and Pegasus region ($f_{\rm
sky}\approx 0.8$ \%) shown in Fig. \ref{fig:MBM_region} we concentrate our
analysis only for this region.

\subsection{Foreground Components in the CMB angular power spectrum}

\begin{figure}
\begin{minipage}[m]{0.5\textwidth}
\rotatebox{0}{\includegraphics[width=1.0\textwidth]{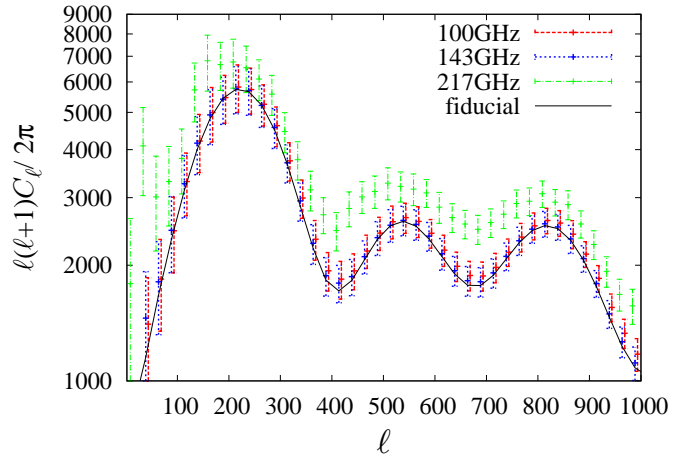}}
\end{minipage}
\caption{The binned power spectra and error bars estimated from $500$
 Monte-Carlo simulations, with foreground components and instrumental
 noises. For clarity, the positions of the bins for $100$ and $217$ GHz are shifted
 by $\Delta\ell=\pm 5$, respectively. The CO components affect the spectrum at small
 angular scales at the $100$ and $217$ GHz
 bands, while the thermal dust component dominates at large angular
 scales at the $217$ GHz band.
 }
 \label{fig:Cls_Foreground}
\end{figure}

We depict the angular power spectra in Fig. \ref{fig:Cls_Foreground}
showing the impact of the foreground components at the MBM  and Pegasus region and
instrumental noises on the power spectra. To estimate the power spectra
we mask all the pixels outside the MBM  and Pegasus region and use the Polspice code
\citep{2001ApJ...561L..11S, 2004MNRAS.350..914C}.  Errors are estimated
by generating five hundred mock CMB and instrumental white noise maps as
described earlier. Because the mask covers most part of the sky ($f_{\rm
sky}\approx 0.8$ \% for the MBM  and Pegasus region) and we thus have large cosmic
variance errors with correlations between neighboring multipoles, we bin
the spectra with the bandwidth $\Delta \ell = 25$.

It has been known that the thermal dust component has larger power at
higher frequencies and affected the spectrum mainly at larger angular scales 
\citep{2000ApJ...530..133T,2004AdSpR..34..483M}.  We find that this holds
true for the MBM  and Pegasus region considered here. On the other hand, we find that
the CO component gives significant contaminations at smaller angular
scales. The contamination can be larger than the instrumental and cosmic
variance errors at $\ell \gtrsim 900$ and $\ell \gtrsim 400$ at the
$100$ and $217$ GHz bands, respectively. Galactic synchrotron emissions
are found 
to be always subdominant in those frequencies and multipole range
because the MBM  and Pegasus region is far away from the galactic disk, and we have
omitted in our current analysis.

\section{Results}
\subsection{Foreground components estimated by the FastICA}
To the sky maps prepared in the previous section we apply the FastICA
algorithm in order to estimate the CO contribution and subtract it from the
maps. As described earlier, we use the sky maps at three frequency bands
to separate out three independent components based on the kurtosis.

In Fig. \ref{fig:sources}, the three sources obtained from the ICA
algorithm are shown. We find that in our deflection algorithm, the
algorithm always finds the CO-like component as the first independent
component ($S^1$), irrespective of the initial condition for the vector
$\bm{w}$. This is caused by the fact that the distribution of the CO
line intensity has the largest non-Gaussianity in terms of the
kurtosis among the three component (CO, dust and CMB). It is also
evident from the figure that the second independent 
component $S^2$ is most responsible for the thermal dust component given
by the dust model of Finkbeiner et al..

\begin{figure}
\rotatebox{0}{\includegraphics[width=0.5\textwidth]{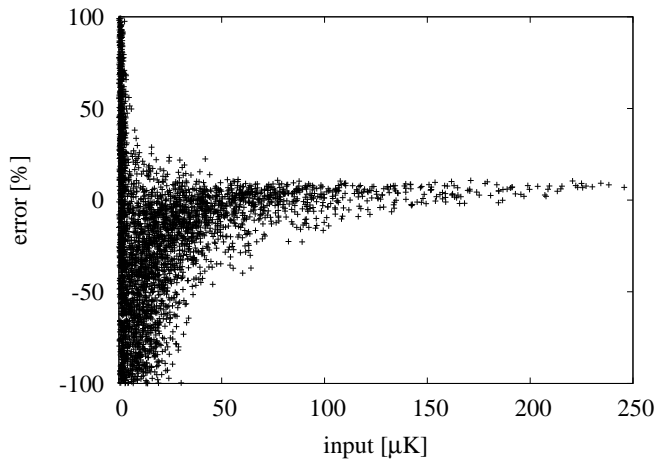}}
\caption{The fractional error in the estimated CO intensity at each
 pixel at the 100GHz band.
 Here the CO is assumed to be the first independent component. While the
 errors
 can be large at pixels in which the CO intensities are intrinsically
 small as $T_{\rm CO} \lesssim 50 \mu{\rm K}$, the errors are less than
 10\% in pixels with large temperature in this particular realization.} 
 \label{fig:scatter}
\end{figure}

\begin{figure}
\begin{minipage}[m]{0.5\textwidth}
\centering\rotatebox{90}{\includegraphics[width=0.8\textwidth]{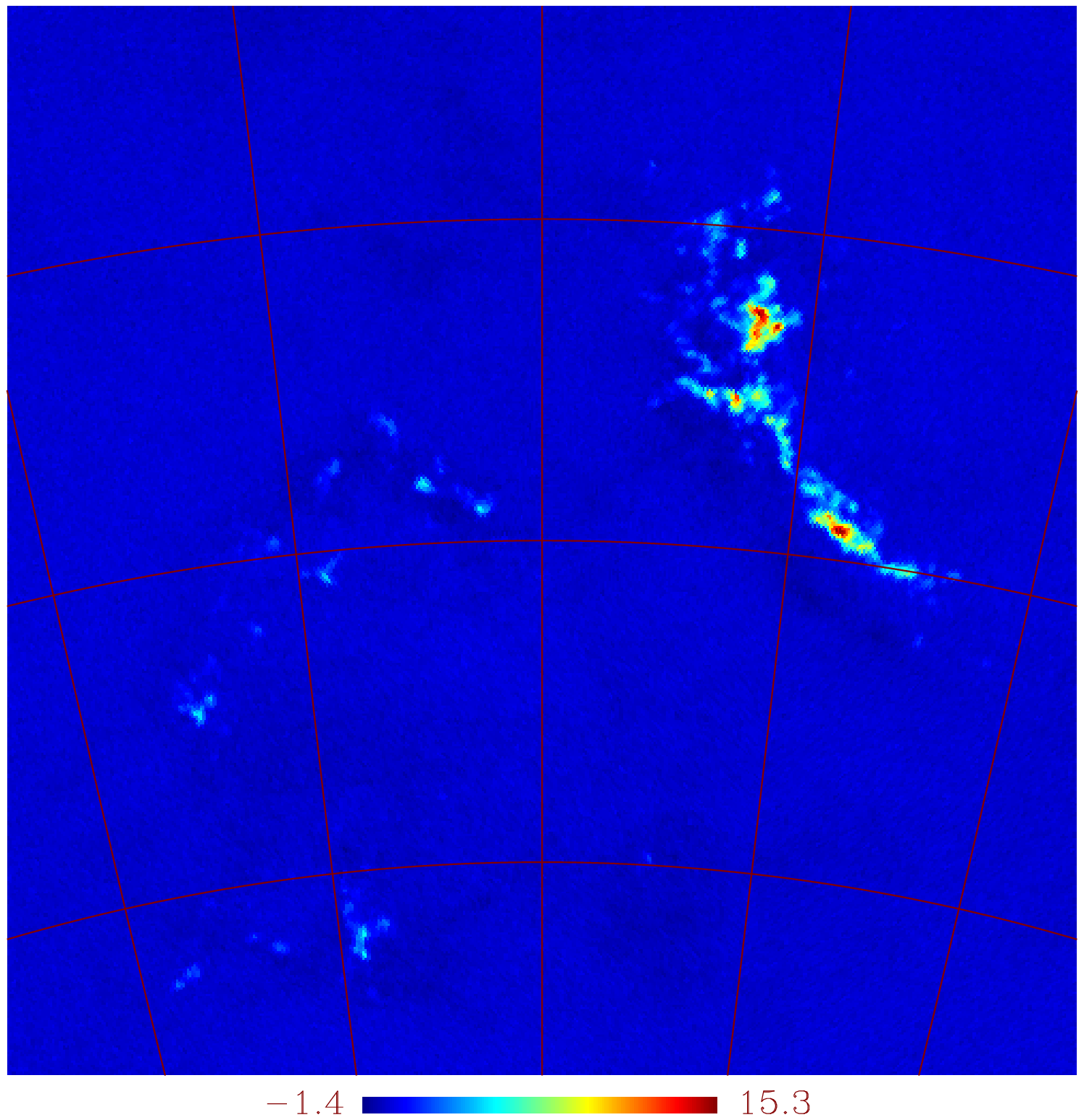}}
\end{minipage}
\begin{minipage}[m]{0.5\textwidth}
\centering\rotatebox{90}{\includegraphics[width=0.8\textwidth]{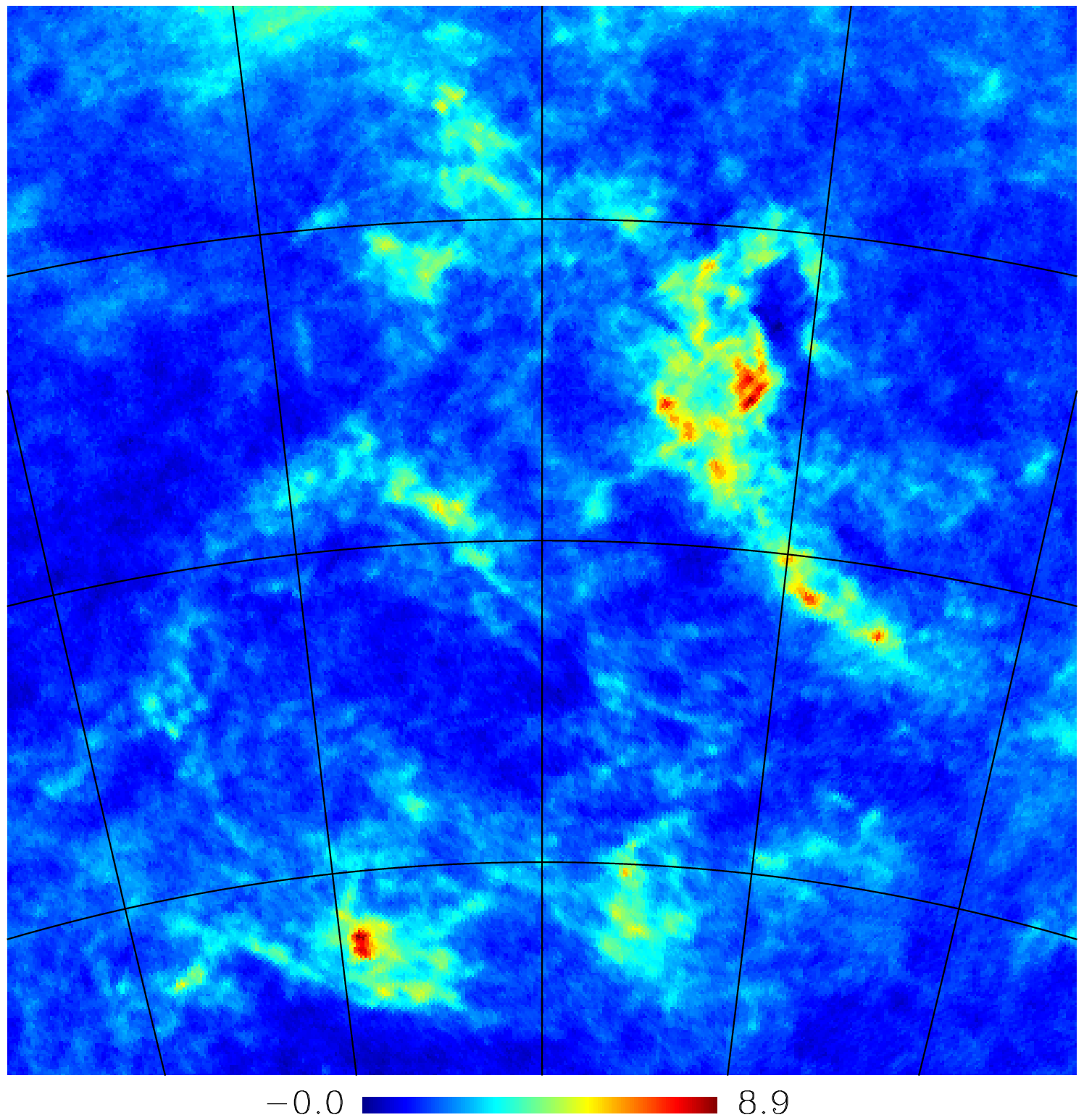}}
\end{minipage}
\begin{minipage}[m]{0.5\textwidth}
\centering\rotatebox{90}{\includegraphics[width=0.8\textwidth]{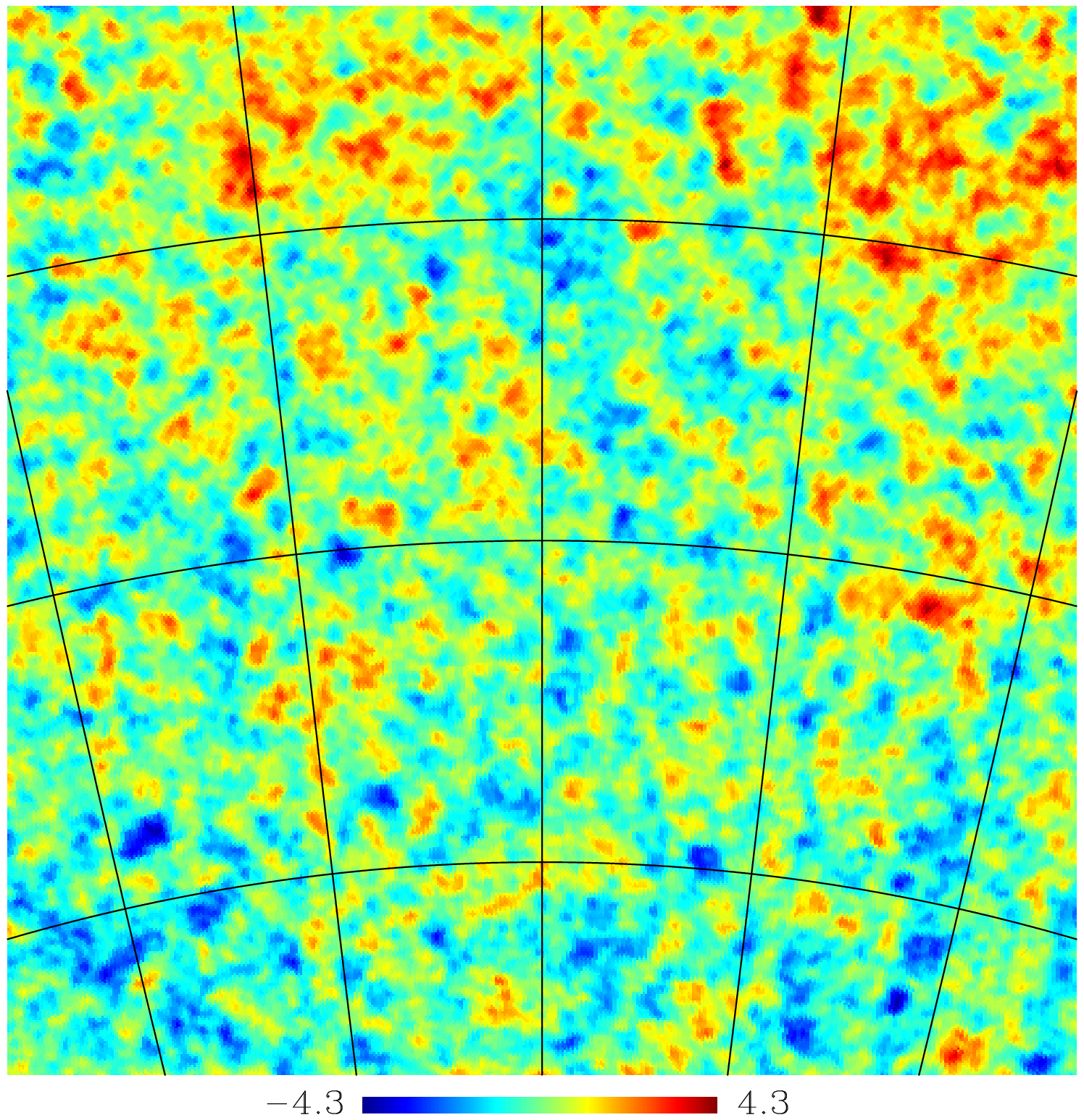}}
\end{minipage}
\caption{Independent components ($S^1, S^2$, and $S^3$) obtained from
 the ICA algorithm. The source with the largest non-Gaussianity is shown
 in 
 the top panel and the smallest in the bottom. It is clear that the ICA
 successfully estimates the CO distribution ($S^1$) in the top
 panel. The second source seems to be thermal dust emissions. The amplitudes are arbitrary because of the
 degeneracy between the sources and mixing matrix components. } \label{fig:sources}
\end{figure}

In order to investigate the performance of the FastICA method as an
estimator of the CO component we make a scatter plot as shown in
Fig. \ref{fig:scatter}. In the figure we show the intensity of the first
independent component estimated by the FastICA at each pixel against
the input CO intensity. We find that the accuracy depends on realizations;
namely, on particular CMB and noise realizations on which the CO
emissions are superimposed. Overall, for most of the realizations the
errors can be less than $\lesssim 30$ \% at pixels where the CO emissions are
strong, while the estimated intensity has large scatter where the CO
intensity is intrinsically small.

Because the performance depends on realizations, we again use the
Monte-Carlo simulations described earlier to rate the performance
statistically as discussed below. 

\subsection{Angular Power Spectrum Estimation}

\begin{figure}
\begin{minipage}[m]{0.5\textwidth}
\rotatebox{0}{\includegraphics[width=1.0\textwidth]{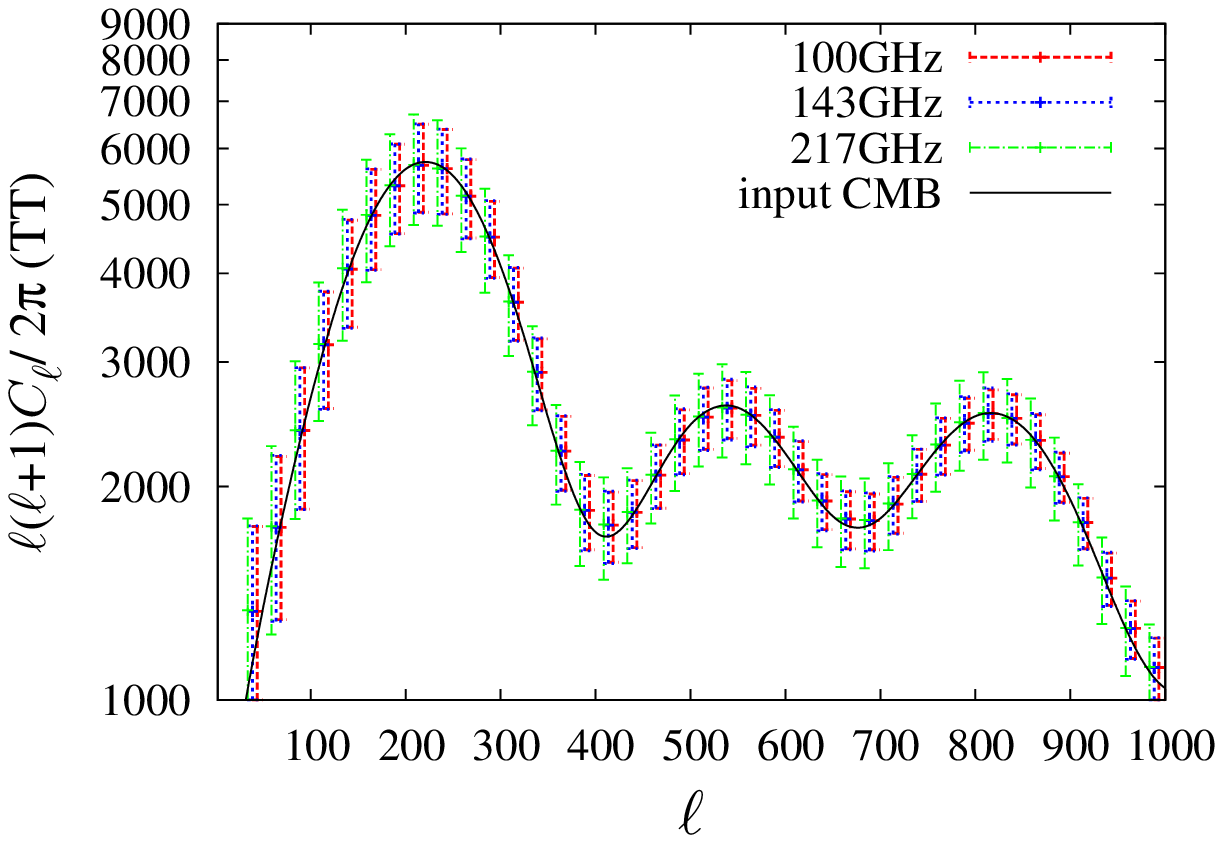}}
\rotatebox{0}{\includegraphics[width=1.0\textwidth]{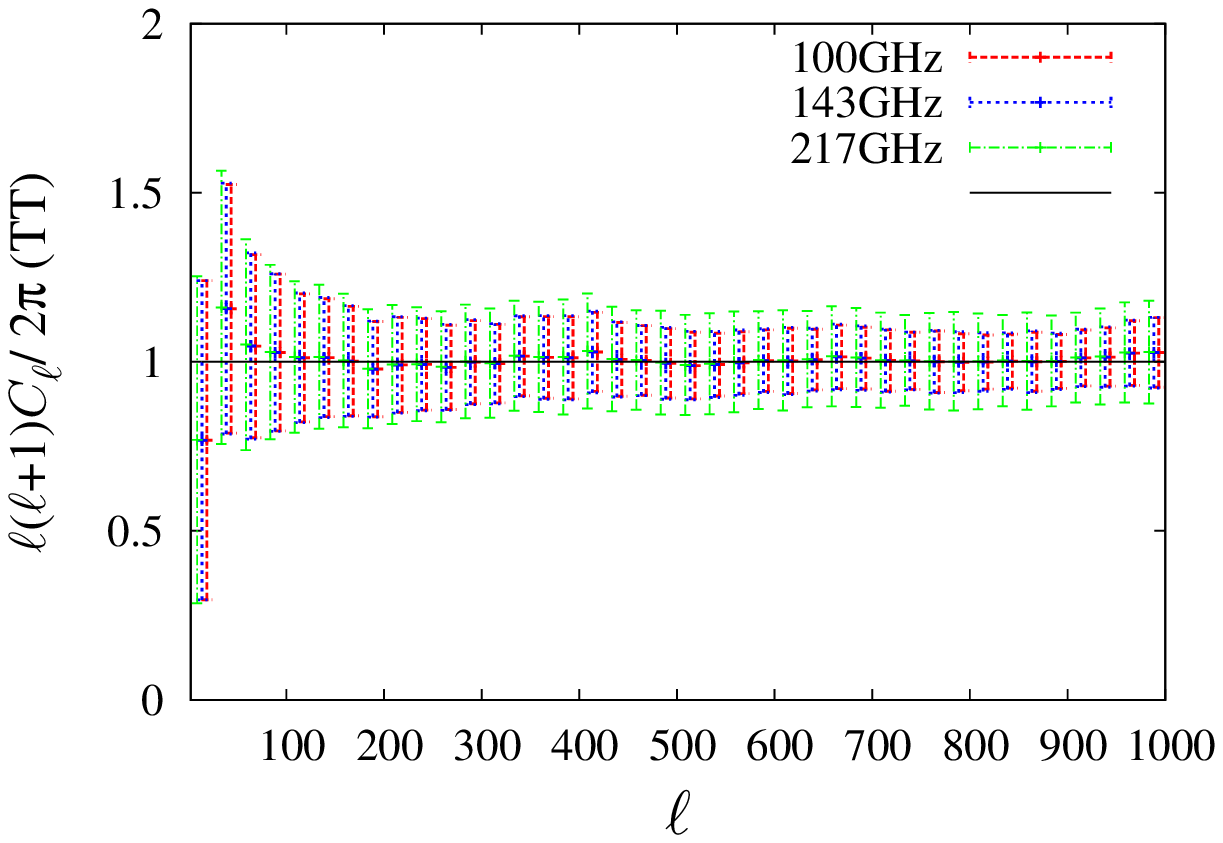}}
\end{minipage}
\caption{The binned power spectrum and error bars estimated from $500$
 Monte-Carlo simulations (top) and fractional errors (bottom). For clarity, the positions of the bins for $100$ and $217$ GHz are shifted
 by $\Delta\ell=\pm 5$, respectively. Acoustic
 structure in the estimated bias is seen; the power is underestimated
 around the acoustic peaks and overestimated around the dips. However, the bias
 is well within the error bars.}
 \label{fig:estimated_Cl}
\end{figure}

To rate statistically the performance of the FastICA method which
depends on realizations, here we apply FastICA method to all the
simulated CMB maps, estimate the foreground components to be removed, and
calculate the CMB angular power spectrum for each realization. The
results are shown in Fig. \ref{fig:estimated_Cl}. In this figure we show
the power spectrum of the CMB component that is estimated as the third
independent component. Clearly, it is found that the additional power
due to the galactic foregrounds found in Fig. \ref{fig:Cls_Foreground}
is successfully removed from all the frequency bands by the ICA, and the
input CMB angular power spectrum is recovered within their error bars.

Interestingly, we find that the error bars are almost unaffected by this
procedure. The errors become larger only about $\lesssim 10$ \% at the
$217$ GHz band where the foreground contamination is the largest, while
at the $100$ and $143$ GHz bands the error bars in the estimations of
the CMB power spectra have almost the same magnitude as those before the
foreground removal.  
The 10\% increase of the errors can be considered as uncertainties in
the FastICA method.

In the bottom panel of Fig. \ref{fig:estimated_Cl} we again depict the
band powers normalized by the input CMB angular power spectrum.
An acoustic-like structure in the estimated bias is seen; the power is
underestimated around the acoustic peaks and overestimated around the
dips. This is partly
because we have binned the power spectrum with an equal weighting in the
$\ell(\ell+1)C_\ell$ space. Note however that the bias is well within
the error bars. 


\section{Summary and Discussion}
In this paper we considered the CMB foreground subtraction
problem, paying particular attention to the lowest two rotational
transitions of CO molecule J=(1-0) and J=(2-1) that contaminate the
PLANCK $100$ GHz and $217$ GHz bands, respectively. Firstly, we estimated
the angular power spectrum of the CO line emissions at the MBM and Pegasus region
observed by NANTEN telescope, and found that the CO line emissions have
significant contribution to the angular power spectrum especially at
small angular scales ($\ell \gtrsim 900$ and $400$ for the $100$ and
$217$ GHz bands, respectively.)

The CO contamination, if it is not to be taken into account correctly,
will cause a wrong estimation of cosmological parameters. In particular,
the parameters related to the primordial fluctuation amplitude such as
the amplitude of the curvature perturbation and its spectral index will
be significantly affected. Indeed, we had found that even for a small
MBM and Pegasus region the bias about the estimations of these parameters are beyond
the $1 \sigma$ error bars, toward larger amplitude and larger spectral
index. We should stress, however,  that this result holds only for the MBM and Pegasus
region of the sky. It will be a future issue how large is the CO
contamination to the estimation of the full sky CMB angular power
spectrum.

Secondly, we applied the FastICA to the Foreground subtraction problem
including the CO line emissions. The FastICA algorithm can separate the
components based on the independency of the components or equivalently
the level of non-Gaussianity, without any prior knowledge of
distribution and frequency dependence of each foreground component.  We
find that CO-like component is extracted as the first independent
component in our deflection algorithm as the CO distribution has the 
largest non-Gaussianity among the components considered here. This fact
can be used to estimate quickly the CO component in the PLANCK data.

Based on the Monte Carlo simulations including CO and thermal dust
emissions as foregrounds, we investigated how the CMB is recovered in
terms of the power spectrum.  Though the accuracy depends on the
particular realization of the instrumental noises and CMB, we found that
the recovery is very well in statistical sense. The success is thanks to
the approximate statistical independence between the foreground
components like CO and background CMB.  This is consistent with the
result in the earlier literature where the authors applied the FastICA
method to the WMAP data and found that it can recover the CMB angular
power spectrum consistent with the spectrum independently derived by the
WMAP team \citep{2007MNRAS.374.1207M}.

\begin{figure}[ht]
\rotatebox{0}{\includegraphics[width=0.5\textwidth]{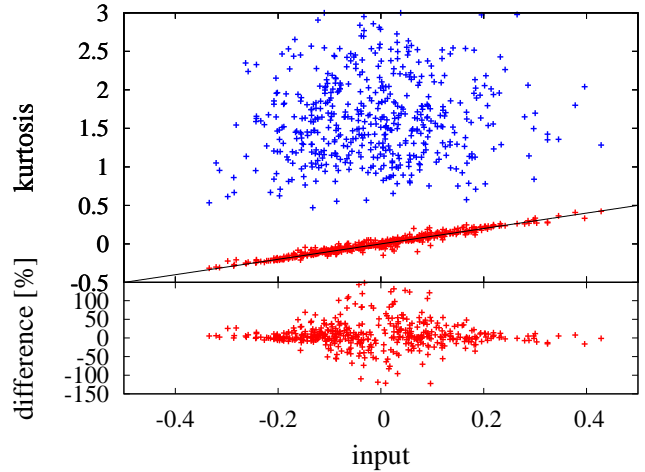}}
\caption{(Top) Kurtoses in the estimated CMB (red) and CMB+Foreground (blue)
 maps at the $217$ GHz band. The straight line represents linear
 relation. (Bottom) Fractional differences between the input and estimated kurtoses.}
 \label{fig:kurtosis}
\end{figure}

Finally, we should comment on the impact of the FastICA method on the
level of non-Gaussianity in the estimated CMB map. Because the method
relies on the non-Gaussianity to estimate the independent components,
naively it should affect the non-Gaussianity of the CMB which probably
has the smallest degree of non-Gaussianity among the components in the
microwave sky. In Fig. \ref{fig:kurtosis}, we show the kurtoses in the
estimated CMB (red) and CMB+Foreground (blue) maps against the value in
the input CMB maps at the $217$ GHz
band in the MBM and Pegasus region. Clearly, it is seen that the bulk of the
kurtosis which comes from the thermal dust and CO components is removed
through the method. 
Interestingly, we find that some portion of the kurtosis in the CMB maps
(which should be zero in the mean in our Gaussian simulation) is
recovered with a scatter about $20$\% when the kurtosis has large value
($|{\rm kurt}|\gtrsim 0.2$).  However, the accuracy depends on
the size of the kurtosis, and the method induces a false signal of the
kurtosis in the 
estimated CMB map when the input kurtosis is too small.
We leave this issue for future investigation.

In conclusion, in this paper we found that the FastICA can efficiently
extract the CO line foregrounds that contaminate the PLANCK HFI
bands. The method will be useful to estimate the CO distribution in
the real PLANCK data, and any foreground component whose distribution is
not known in advance in the future CMB experiments.

\acknowledgements
One of the authors (KI) would like to thank T. Matsumura, E. Komatsu,
 and O. Dor\'{e} for helpful comments and useful discussions.  KI also
 thanks A. Asai for her kind support on the IDL manipulation.  This work
 has been supported in part by Grant-in-Aid for Scientific Research
 Nos. 24340048 (KI), 22740119
 (HY), 23340046, 24111707 (TTT), and 24224006 (YF) from the Ministry of
 Education, Sports, Science and Technology (MEXT) of Japan, and by Grant-in-Aid for the Global Center of Excellence program
 at Nagoya University "Quest for Fundamental Principles in the Universe:
 from Particles to the Solar System and the Cosmos" from the MEXT of
 Japan. The work has also been supported by the Strategic Young Researcher
 Overseas Visits Program for Accelerating Brain Circulation Nos. R2405
 (TTT) and R2211 (YF) commissioned by JSPS.

\bibliographystyle{hapj} 
\bibliography{ms} 

\end{document}